\documentclass[twocolumn]{aastex631}
\usepackage[utf8]{inputenc}
\usepackage{multirow}
\usepackage{float}
\usepackage{natbib}
\usepackage[]{threeparttable}
\usepackage{lineno}
\usepackage{epstopdf}
\newcommand{\hi}{H\,{\sc i}}
\newcommand{\hii}{H\,{\sc ii}}
\newcommand\kms{km$\,$s$^{-1}$}

\shortauthors{Mutlu-Pakdil et al.}

\begin{document}

\title{Hubble Space Telescope Imaging of Three Isolated Faint Dwarf Galaxies Beyond the Local Group: Pavo, Corvus~A, and Kamino}

\correspondingauthor{B. Mutlu-Pakdil}
\email{Burcin.Mutlu-Pakdil@dartmouth.edu}

\author[0000-0001-9649-4815]{Bur\c{c}in Mutlu-Pakdil}
\affil{Department of Physics and Astronomy, Dartmouth College, Hanover, NH 03755, USA}

\author[0000-0002-5434-4904]{Michael G. Jones}
\affil{Steward Observatory, University of Arizona, 933 North Cherry Avenue, Tucson, AZ 85721, USA}

\author[0000-0003-4102-380X]{David J. Sand}
\affil{Steward Observatory, University of Arizona, 933 North Cherry Avenue, Tucson, AZ 85721, USA}

\author[0000-0002-1763-4128]{Denija Crnojevi\'{c}}
\affil{University of Tampa, 401 West Kennedy Boulevard, Tampa, FL 33606, USA}

\author[0000-0003-4394-7491]{Kai Herron}
\affil{Department of Physics and Astronomy, Dartmouth College, Hanover, NH 03755, USA}

\author[0000-0002-1468-9668]{Jay Strader}
\affiliation{Center for Data Intensive and Time Domain Astronomy, Department of Physics and Astronomy,\\ Michigan State University, East Lansing, MI 48824, USA}

\author[0000-0002-5177-727X]{Dennis Zaritsky}
\affil{Steward Observatory, University of Arizona, 933 North Cherry Avenue, Tucson, AZ 85721, USA}

\author[0000-0001-8354-7279]{Paul Bennet}
\affiliation{Space Telescope Science Institute, 3700 San Martin Drive, Baltimore, MD 21218, USA}

\author[0000-0001-8251-933X]{Alex Drlica-Wagner}
\affil{Astronomy \& Astrophysics, University of Chicago, Chicago, IL 60637 USA}
\affil{Fermi National Accelerator Laboratory, Batavia, IL, USA}

\author[0009-0005-9002-4800]{Quinn O. Casey}
\affiliation{Department of Physics and Astronomy, Dartmouth College, Hanover, NH 03755, USA}

\author[0000-0001-9775-9029]{Amandine~Doliva-Dolinsky}
\affil{Department of Physics, University of Surrey, Guildford GU2 7XH, UK}

\author[0000-0001-7618-8212]{Richard Donnerstein}
\affil{Steward Observatory, University of Arizona, 933 North Cherry Avenue, Tucson, AZ 85721, USA}

\author[0000-0001-8245-779X]{Catherine E. Fielder}
\affil{Steward Observatory, University of Arizona, 933 North Cherry Avenue, Tucson, AZ 85721, USA}

\author[0000-0001-5368-3632]{Laura C. Hunter}
\affiliation{Department of Physics and Astronomy, Dartmouth College, 6127 Wilder Laboratory, Hanover, NH 03755, USA}

\author[0000-0002-8040-6785]{Annika H. G. Peter}
\affiliation{CCAPP, Department of Physics, and Department of Astronomy, The Ohio State University, Columbus, OH 43210, USA}

\author[0000-0002-8217-5626]{Deepthi S. Prabhu}
\affil{Steward Observatory, University of Arizona, 933 North Cherry Avenue, Tucson, AZ 85721, USA}

\author[0000-0002-0956-7949]{Kristine Spekkens}
\affil{Department of Physics, Engineering Physics and Astronomy, Queen's University, Kingston, ON K7L 3N6, Canada}
 
\begin{abstract}

We present new {\it Hubble Space Telescope} ({\it HST}) imaging of three recently discovered star-forming dwarf galaxies beyond the Local Group: Pavo, Corvus~A, and Kamino. The discovery of Kamino is reported here for the first time. They rank among the most isolated faint dwarf galaxies known, hence they provide unique opportunities to study galaxy evolution at the smallest scales, free from environmental effects of more massive galaxies. Our {\it HST} data reach $\sim$2-4~magnitudes below the tip of the red giant branch for each dwarf, allowing us to measure their distances, structural properties, and recent star formation histories (SFHs).  All three galaxies contain a complex stellar population of young and old stars, and are typical of field galaxies in this mass regime ($M_V=-10.62\pm0.08$ and $D=2.16^{+0.08}_{-0.07}$~Mpc for Pavo, $M_V=-10.91\pm0.10$ and $D=3.34\pm0.11$~Mpc for Corvus~A, and $M_V=-12.02\pm0.12$ and $D=6.50^{+0.15}_{-0.11}$~Mpc for Kamino). Our {\it HST}-derived SFHs reveal differences among the three dwarfs: Pavo and Kamino show relatively steady, continuous star formation, while Corvus~A formed $\sim$60\% of its stellar mass by 10~Gyr ago. These results align with theoretical predictions of diverse evolutionary pathways for isolated low-mass galaxies.

\end{abstract}

\section{Introduction} \label{sec:intro}

Faint dwarf galaxies are unique laboratories to test the predictions of cosmological models on small scales \citep{Bullock17,Sales2022}. The Local Group dwarf satellites are often used for a testing ground due to the detail with which they can be studied \citep[e.g.,][]{Weisz2014,Brown2014,Weisz2015,Skillman2017,Garrison-Kimmel2019,Nadler2021,Akins2021,Applebaum2021,Manwadkar2022,Santos2024,Dublin2025,Amandine2025}. However, relying exclusively on them risks `over-tailoring' the models to a single environment.

After infall, the Local Group satellite dwarfs evolve in the hot halo gas of the Milky Way (MW) and M31, and are heavily affected by this environment \citep{Spekkens14,Putman21}. While the least luminous satellites of the MW and M31 are almost certainly quenched by reionization \citep{Brown2014,Rodriguez-Wimberly2019,Rey2020}, the “classical” dwarf galaxy satellites -- the best-studied galaxies in the dwarf galaxy regime -- are driven by a combination of factors -- including reionization, feedback, ram pressure, and tidal interactions \citep{Mayer2006,Akins2021,Samuel2022,Mao2024,Pathak2025}.  These physical processes can act simultaneously, and disentangling the relative impact of each is challenging. To better associate observed differences with specific physical mechanisms, it is therefore essential to study isolated dwarf galaxies, where some environmental effects, like ram pressure and tidal stripping, are largely absent (but see \citealt{Wright2019,Benavides2025}). These isolated low-mass dwarfs provide cleaner and more robust laboratories for testing dark matter halo structure and cosmological processes (i.e., cosmic reionization).

Identifying isolated dwarf galaxies poses a formidable challenge due to their inherent faintness ($M_{V}\gtrsim-12$). Unlike satellite dwarfs, which can be found in proximity to a host galaxy, field dwarfs present a much more elusive quest, as we lack clear guidance on where to look for them. This inherent uncertainty significantly compounds the difficulty of their discovery. Until recently, Leo~P was the only known isolated star-forming galaxy with $M_{\star} \lesssim 10^6~M_{\odot}$. It was discovered through a blind \hi~Arecibo Legacy Fast ALFA (ALFALFA) survey \citep{Giovanelli2013}, but subsequent attempts to find similar objects have been unsuccessful. Leo~P has only $\sim 3 \times 10^5 M_{\odot}$ in stars and a comparable amount of gas \citep{McQuinn2024}. It resides in an exceptional regime where it is only just massive enough to retain its gas through cosmic reionization and subsequent episodes of star formation \citep[e.g.,][]{Rey2020,Gutcke2022}. It is outside the Local Group but close enough at 1.6~Mpc to be studied with resolved stars (allowing in-depth investigations). However, with only one system, it is difficult to draw broad conclusions about low-mass star-forming galaxies evolving in isolation.  

\begin{figure*}[!ht]
\centering
\includegraphics[width = 0.323\linewidth]{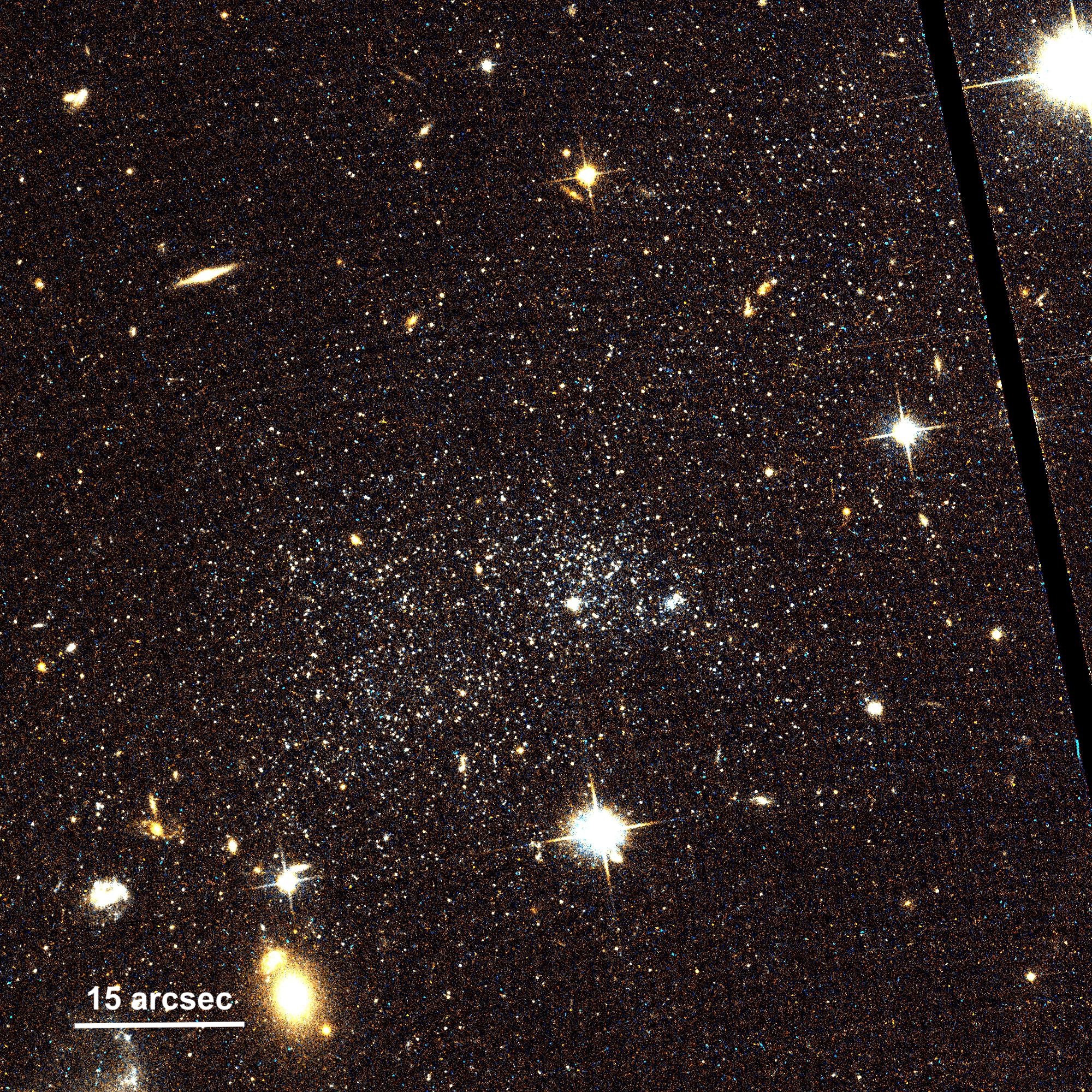}
\includegraphics[width = 0.323\linewidth]{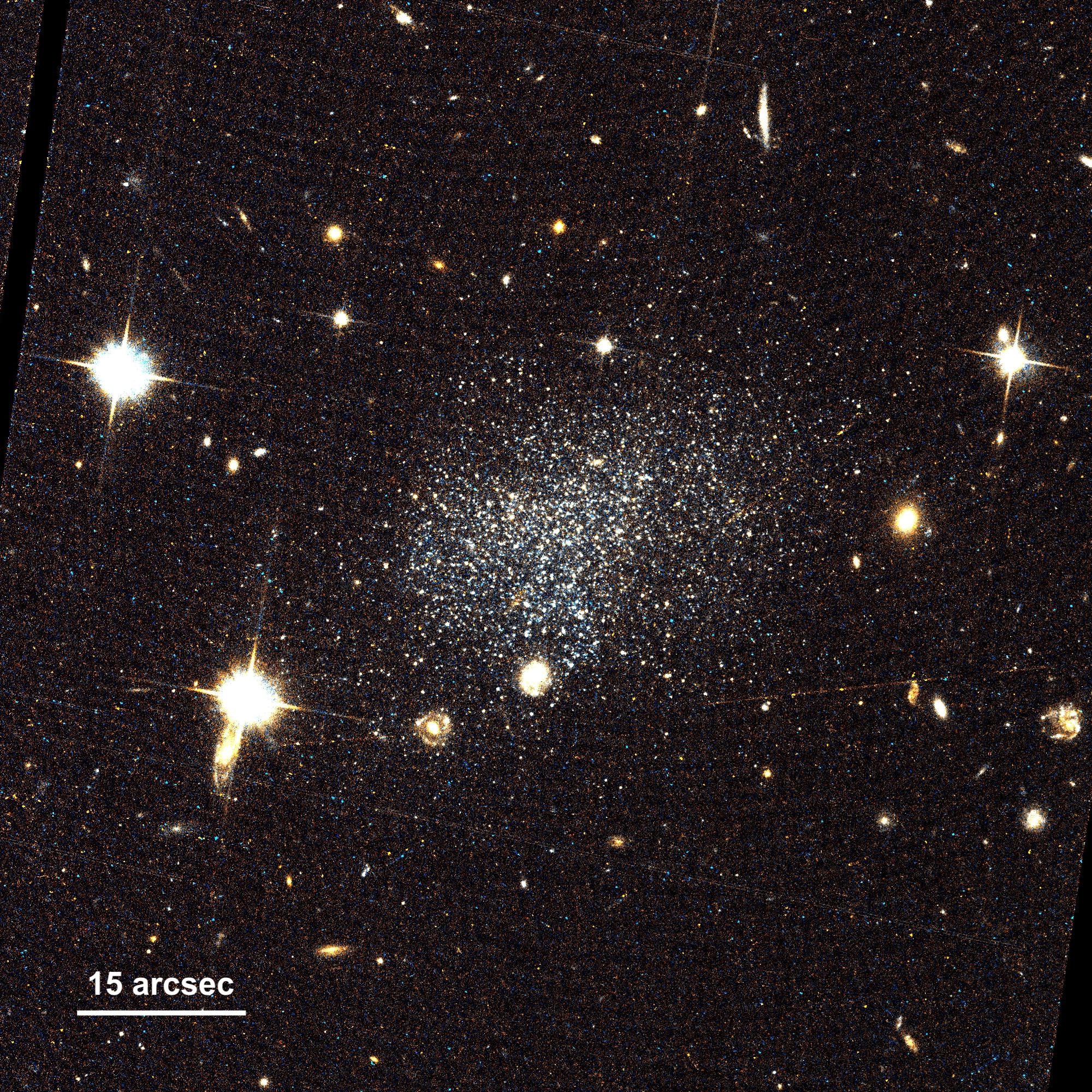}
\includegraphics[width = 0.323\linewidth]{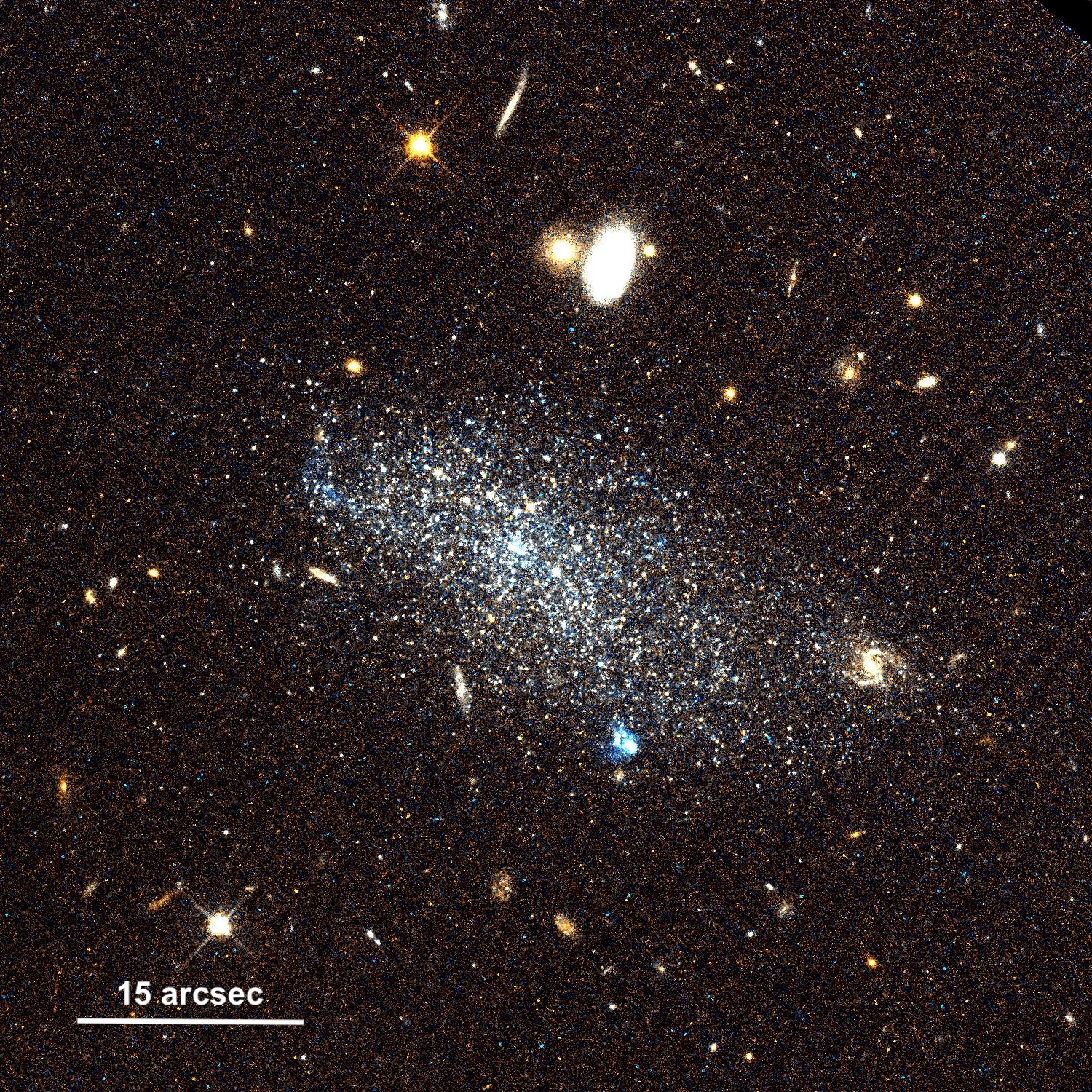}
\caption{Three-color {\it HST}/ACS images of Pavo (left), Corvus~A (middle), and Kamino (right). The images are created by combining the F606W band image (Blue), the average of F606W and F814W band images (Green), and the F814W band image (Red).
North is up, and east is to the left. \label{fig: image}}
\end{figure*} 

Our new survey SEAMLESS (the SEmi-Automated Machine LEarning Search for Semi-resolved galaxies) is designed to deliver a statistical census of the faintest dwarfs beyond the Local Group across diverse environments, including elusive isolated dwarf galaxies, providing a rare opportunity to assess the impact of environment and reionization on the lowest-mass systems \citep{Jones23}. Our search has revealed numerous nearby, low-mass galaxy candidates in the Data Release 10 of the DESI Legacy Imaging Surveys. These include Pavo \citep{Jones23} and Corvus~A \citep{Jones24}, which have since been confirmed, as well as Kamino, whose discovery we present here for the first time. Kamino was also independently identified during the Dark Energy Survey Year~6 low-surface-brightness galaxy search (K. Herron et al., in prep). In ground-based imaging, all three appear isolated, with no larger neighboring galaxy in the projected field. As it is very difficult to find nearby isolated star-forming dwarfs, it is important to thoroughly characterize Pavo, Corvus~A, and Kamino, and gain a deeper understanding of their nature.

In this work, we present {\it HST} observations of Pavo, Corvus~A, and Kamino (see Figure~\ref{fig: image}). In Section~\ref{sec:observations}, we describe the details of our {\it HST} data and photometry.  In Section~\ref{sec:prop}, we measure the properties of our dwarfs, including their stellar populations, distance, structural parameters, gas content, and star formation histories. Finally, we compare our results with those of other isolated dwarf galaxies and Local Group satellites of similar brightness, and summarize our findings in Section~\ref{sec:conclusion}.

\setcounter{table}{0}
\begin{table}[h!]
\renewcommand{\thetable}{\arabic{table}}
\centering
\caption{Summary of {\it HST} Observations and Completeness.} \label{tab:obs}
\begin{tabular}{lcccccc}
\tablewidth{0pt}
\hline
\hline
Name & UT Date &  Filter & Exp  & 50\%  & 90\% \\
{}         & {}      &  {}     & (s)  & (mag) & (mag) \\
\hline
Pavo   &  2024 May 16  &  $F606W$ & 1006 & 27.47 & 26.50    \\
{}     &  2024 May 16  &  $F814W$ & 1268 & 26.81 & 26.12    \\
Corvus~A &  2024 April 19 &  $F606W$ & 980 & 27.49 & 26.53    \\
{}     &  2024 April 19 &  $F814W$ & 1027 & 26.82 & 26.11    \\
Kamino &  2024 October 5 &  $F606W$ & 995 & 27.46 & 26.45    \\
{}     &  2024 October 5 &  $F814W$ & 995 & 26.75 & 26.13    \\
\hline
\end{tabular}
\end{table}

\section{{\it HST} Observations and Photometry} \label{sec:observations}

We obtained {\it HST} follow-up observations of Pavo, Corvus~A, and Kamino (GO-17514, PI: Mutlu-Pakdil) with the Wide Field Channel (WFC) of the Advanced Camera for Surveys (ACS). Each target was observed for a total of one orbit in the F606W and F814W filters (see Table~\ref{tab:obs}). Coordinated parallel observations\footnote{Due to the reduced gyro mode, we could not obtain coordinated parallel observations of Kamino.} of Pavo and Corvus~A were simultaneously obtained with the Wide Field Camera 3 (WFC3) UVIS channel, with the same filters. These parallel pointings serve as a control field to assess the impact of foreground star/background galaxy contamination.

We perform PSF photometry on pipeline-produced FLC images with the latest version (2.0) of DOLPHOT \citep{Dolphin2000}. We follow the recommended preprocessing steps and use the suggested input parameters from the DOLPHOT User Guide. The initial photometry is culled with the following criteria: the sum of the crowding parameters in the two bands is $<1$, the squared sum of the sharpness parameters in the two bands is $<0.1$, the signal-to-noise ratio is $>4$, and the object type is $\leq2$ in each band. We correct for MW extinction using the \citet{Schlegel98} reddening maps with the coefficients from \citet{Schlafly11}. Extinction-corrected photometry is used throughout this work. 

We perform artificial star tests to assess the photometric uncertainties and completeness of our observations. A total of $\sim$1,000,000 stars were added one at a time using the artificial star utilities in DOLPHOT, uniformly distributed both in color-magnitude space (i.e., $20$$\leq$F606W$\leq$$30$ and $-0.5$$\leq$F606W$-$F814W $\leq$$1.5$) and spatially across the field of view. Photometry and quality cuts were carried out in the same way as those applied to the original photometry. Our {\it HST} data are 50\% (90\%) complete at F606W $\sim$27.5 (26.5)~mag and F814W $\sim$26.8 (26.1)~mag (see Table~\ref{tab:obs}).

\section{Properties of Pavo, Corvus~A, and Kamino \label{sec:prop}}
\subsection{Color-Magnitude Diagram \label{sec:cmd}}

\begin{figure*}
\centering
\includegraphics[width = 0.65\linewidth]{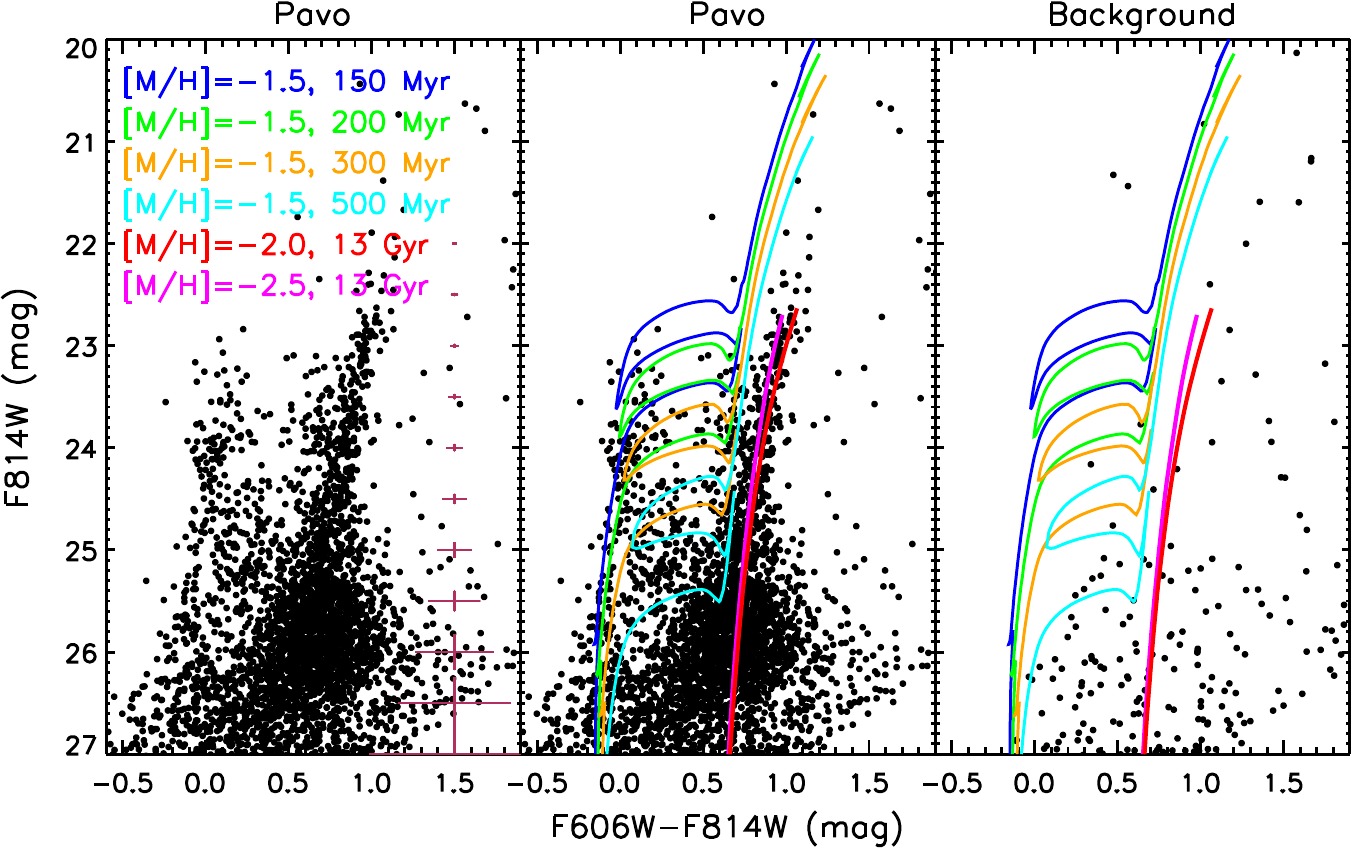}
\includegraphics[width = 0.65\linewidth]{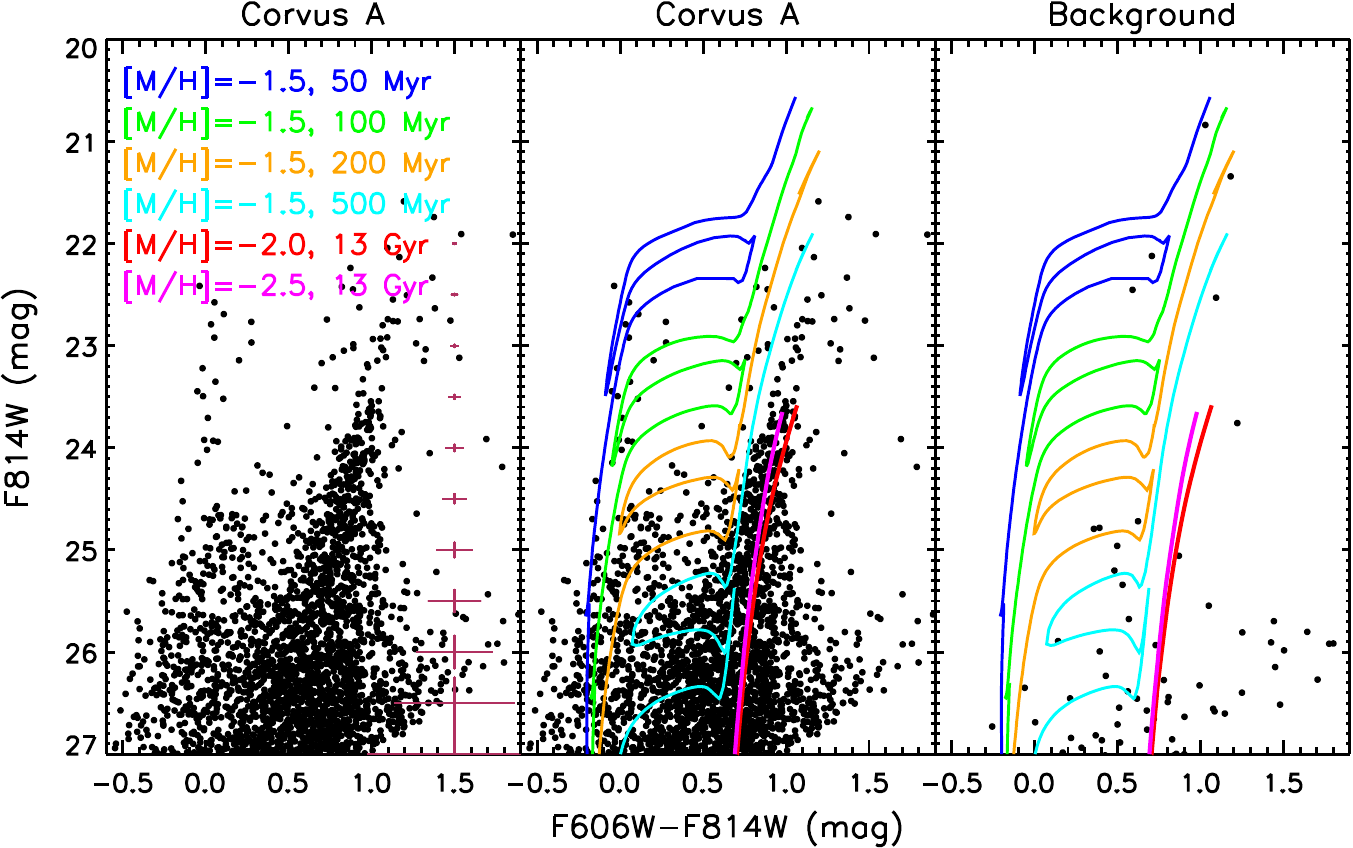}
\includegraphics[width = 0.65\linewidth]{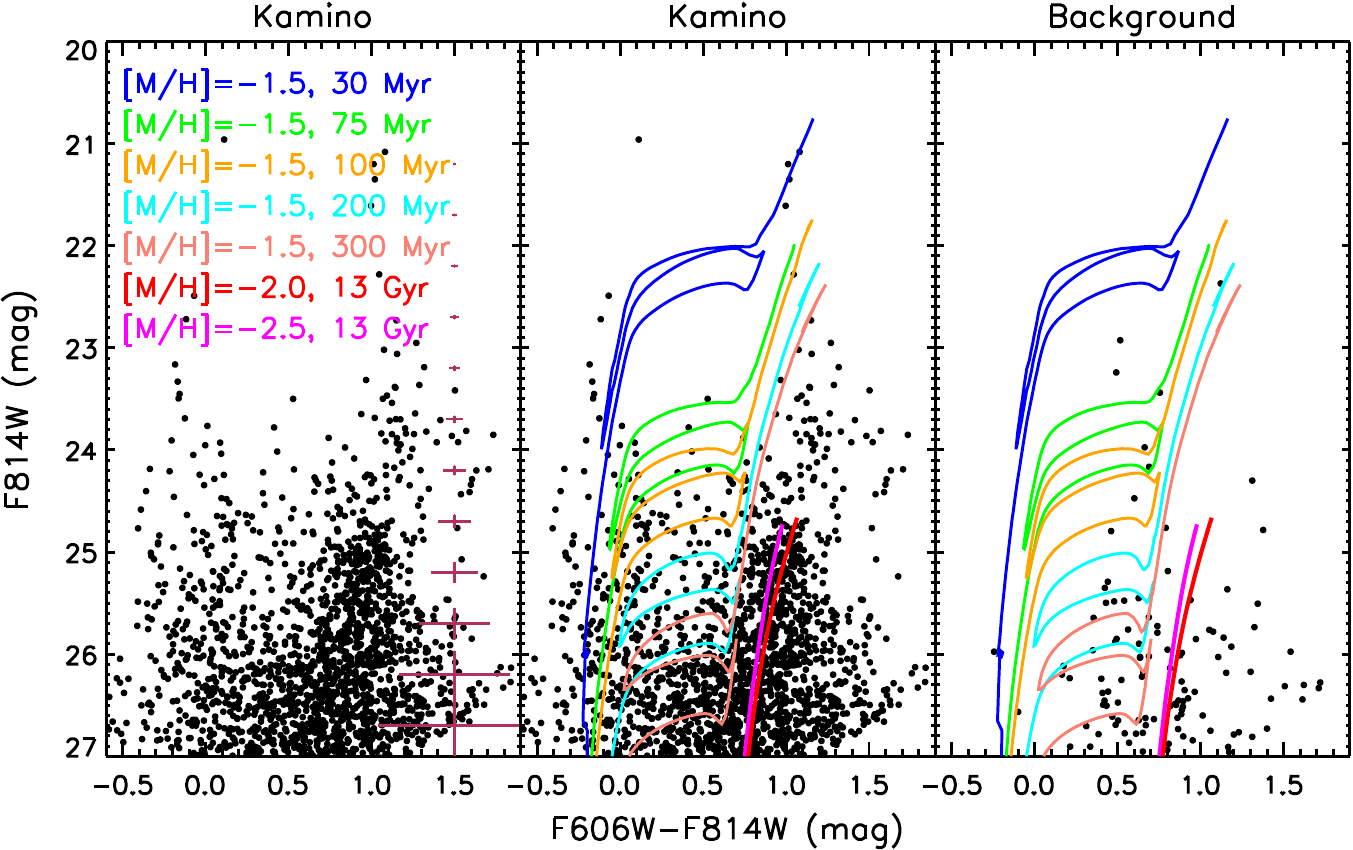}
\caption{{\it HST} CMDs of Pavo (top), Corvus~A (middle), and Kamino (bottom). Left: {\it HST} CMDs showing the stars within $\sim$2$\times$$r_h$ of each dwarf galaxy. Maroon error bars show the mean photometric errors from artificial stars. Center: the same as in the left panel, with PARSEC isochrones for different RGB stellar population ages overlaid. Right: CMD of a representative field region of equal area far away from each dwarf. Due to Pavo's larger size, the WFC3 data are used to provide a control field to assess the impact of foreground star/background galaxy contamination. \label{fig:cmd} }
\end{figure*}

Figure~\ref{fig:cmd} shows the {\it HST} CMDs, which include stars within two half-light radii (2$\times$$r_h$, see Table~\ref{tab:dwarfs} and Section~\ref{sec:str}). Overplotted on the CMDs as red and magenta lines are the PARSEC isochrones \citep{Bressan2012} for 13~Gyr and [M/H] = -2.5 and -2.0 dex, respectively. Maroon error bars show the mean photometric errors from artificial stars, and are plotted at an arbitrary color for convenience. Each dwarf is clearly resolved into its constituent red giant branch (RGB) stars in the {\it HST} data, and shows an old, metal-poor stellar population. 
They all also contain several blue stars with F606W$-$F814W $<0.5$. The center panels show the CMDs with the isochrones for different stellar population ages overlaid. The metallicity for the younger populations is chosen somewhat arbitrarily to be higher than that of the older stellar populations. The blue stars are consistent with young stellar populations, with ages ranging between 150-500~Myr for Pavo, 50-500~Myr for Corvus~A, and 30-300~Myr for Kamino. The right panels show a CMD of a field region of equal size located far from the dwarf. We present background regions within the ACS field for Corvus~A and Kamino, but we use a comparable background region from the WFC3 coordinated parallel field for Pavo due to its larger size. These CMDs serve as a typical field area in the vicinity of the dwarf, helping to assess field contamination. We present their detailed SFH measurements in Section~\ref{sec:sfh}. 

\begin{figure}
\centering
\includegraphics[width = 0.83\linewidth]{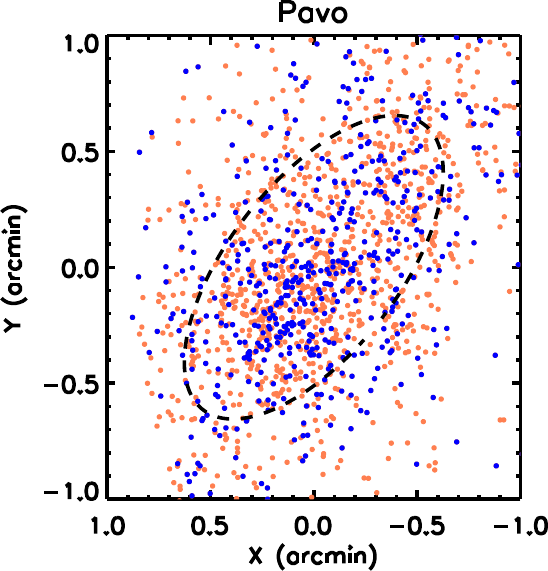}
\includegraphics[width = 0.8\linewidth]{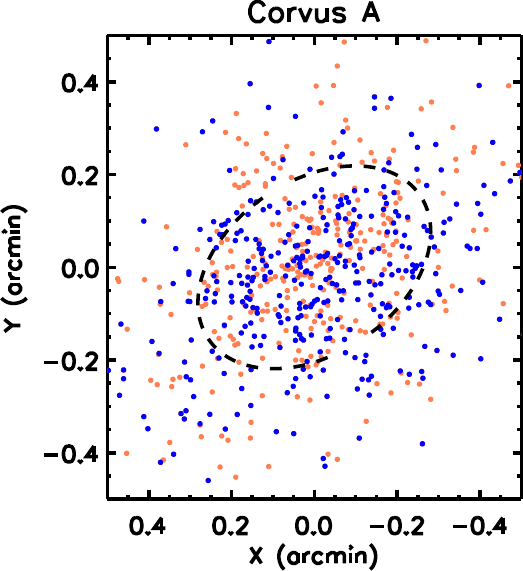}
\includegraphics[width = 0.8\linewidth]{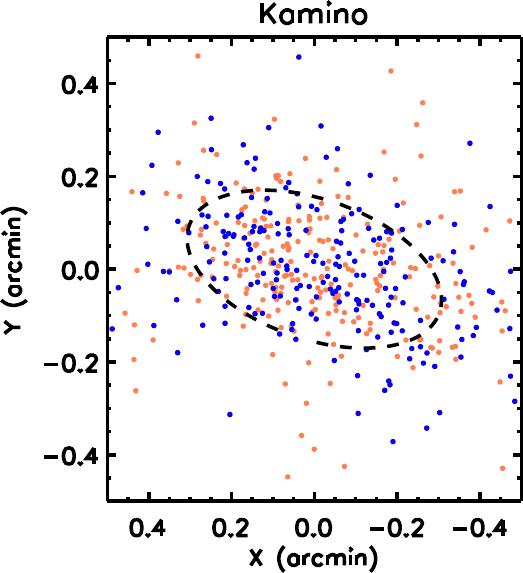}
\caption{Spatial distribution of young and old stars in our dwarfs as defined in the text (RGB in coral, young stars in blue). The dashed ellipse represents one half-light radius. Blue young stars in Pavo display a relatively clumpy and irregular distribution, offset from the center.  \label{fig:position}}
\end{figure}

Figure~\ref{fig:position} shows the spatial distribution of stars, categorized into RGB stars (coral) and young blue stars (blue), based on their position in the CMD. Young stars are selected from those that are consistent with the younger isochrones (i.e., F814W $< 26.0$~mag and F606W$-$F814W $< 0.5$~mag). RGB stars are the same ones we use to derive the structural parameters presented in Section~\ref{sec:str}. The black dashed line represents the half-light radius of each dwarf. RGB stars present a regular and elongated shape, with no clear signs of disturbance. Blue stars in Pavo display a relatively clumpy and irregular distribution, offset from the center. This is consistent with the deep ground-based follow-up observations of Pavo that we reported in \citet{Jones23}.

\subsection{Distance \label{sec:dist}}

We measure distances to our targets using the TRGB method \citep[e.g.,][]{Lee1993,Salaris2002,Rizzi2007}, as described in \citet{Crnojevic19} and \citet{BMP2022}. We first apply a color correction to our photometry, to reduce the curvature and width of the RGB and increase the sharpness in the break of the luminosity function, following \cite{jang17} (their formula~5 and Table~6); we then compute the observed luminosity function for RGB stars, applying a color cut of F606W$-$F814W$>0.6$ to avoid any contamination from possible young populations (see Figure~\ref{fig:cmd}). The luminosity function is fit with a model that is convolved by the appropriate photometric uncertainty and completeness function as derived from our artificial star tests. Our final uncertainties combine the fitting uncertainties (which include the artificial star test results), the uncertainties from the TRGB zeropoint calibration and the applied color correction, and an assumed 10\% uncertainty on the adopted extinction value, added in quadrature. The TRGB values, the distance moduli, and the distances for our targets are reported in Table~\ref{tab:dwarfs}. 

\subsection{Structural Properties \label{sec:str}}

We derive structural parameters of our targets using the maximum-likelihood (ML) method described by \citet{Sand12}, which builds on the approach of \citet{Martin08}. We use stars consistent with an old, metal-poor isochrone in color-magnitude space after taking into account photometric uncertainties, within our 90\% completeness limit. When photometric errors are less than 0.1~mag, we inflate the uncertainty to 0.1~mag to select stars for our ML analysis. We fit a standard exponential profile plus a constant background to the data. The resulting structural parameters are summarized in Table~\ref{tab:dwarfs}. We determine uncertainties by bootstrapping the data 1,500 times and recalculating the structural parameters for each sample.

Leo~P is the prototypical example of a low-mass, isolated, star-forming dwarf galaxy in the nearby universe, and provides a valuable benchmark for comparison with our targets. To enable a direct comparison, we analyze the archival {\it HST} data of Leo~P (GO 13376; PI: McQuinn) by performing PSF photometry following the same procedure described in Section~\ref{sec:observations}, and apply our ML analysis to derive its structural parameters (see Table~\ref{tab:dwarfs}). Our results for Leo~P are consistent with previously reported values \citep{McQuinn2013,McQuinn2015}, and rederiving them using our methodology ensures a consistent and reliable comparison with our sample (see Section~\ref{sec:conclusion}). 

\subsection{Star Formation Histories \label{sec:sfh}}

To fit the SFHs, we follow the established methodology of StarFISH \citep{Harris+2001}, but have re-implemented the steps in Python rather than the original FORTRAN. Here we give an overview of the fitting process, which is described in more detail in \citet{Mike25_sfh}. StarFISH and other CMD-based SFH fitting codes \citep[e.g.,][]{MATCH,IAC-pop,Garling+2024} fit observed CMDs with linear combinations of artificial single stellar population (SSP) CMDs spanning a range of ages and metallicities (and sometimes distances). The combined weights of these SSPs equate to an SFH. 

\begin{figure*}[!t]
    \centering
    \includegraphics[width=0.32\linewidth]{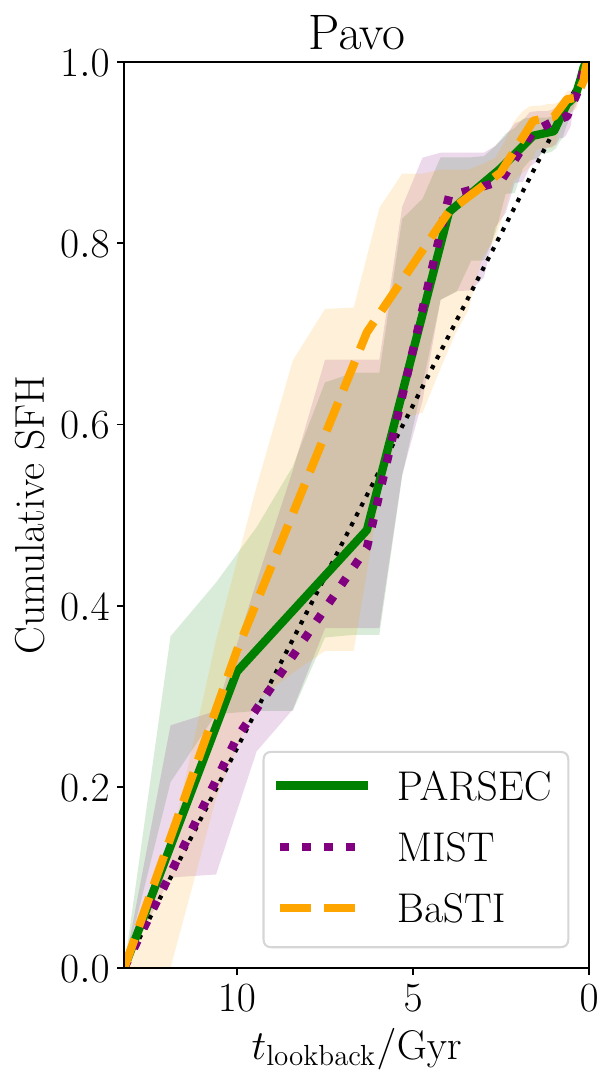}
    \includegraphics[width=0.32\linewidth]{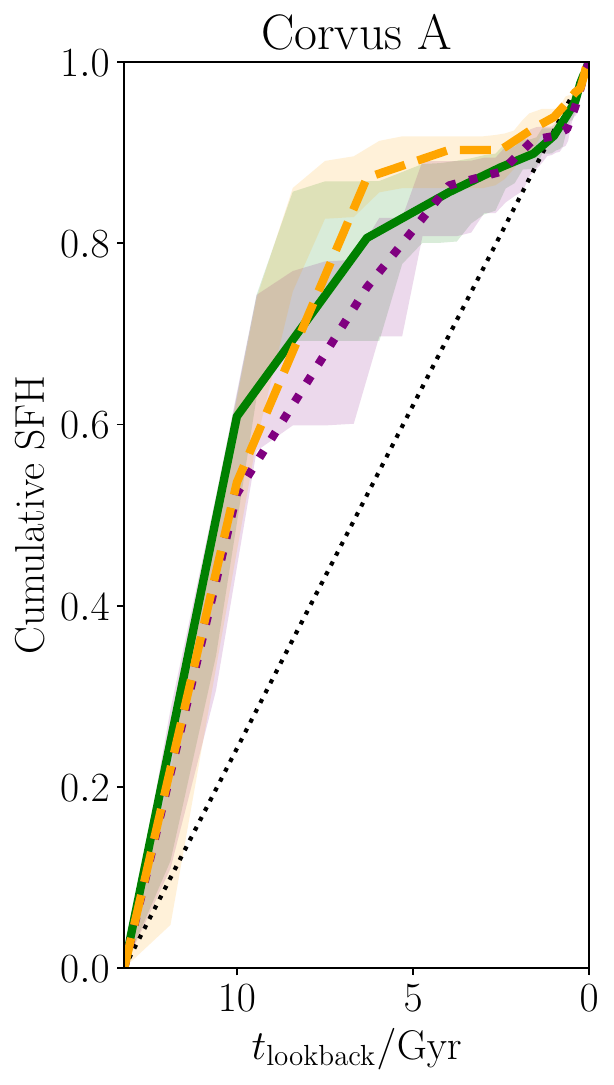}
    \includegraphics[width=0.32\linewidth]{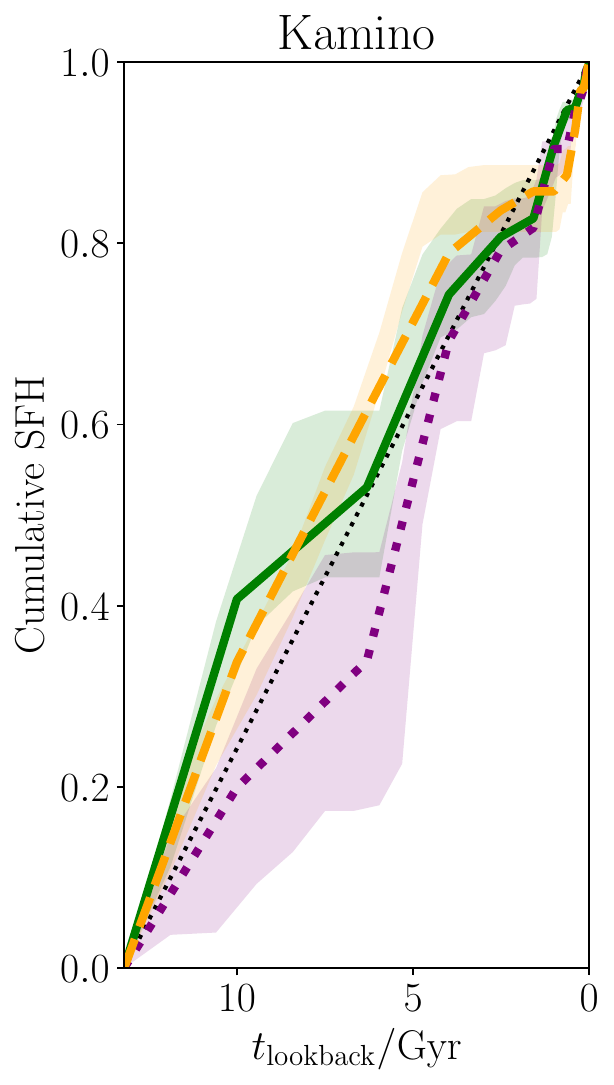}
    \caption{Cumulative star formation histories of Pavo (left), Corvus~A (center), and Kamino (right). The fits from the three different isochrone libraries are shown with different colors and line styles. The shaded color bands indicate the 1-$\sigma$ uncertainties of each model based on re-fitting Monte Carlo realizations of the best-fit artificial CMD. The diagonal dashed black lines represent a constant SFR for the age of the universe.}
    \label{fig:cumSFHs}
\end{figure*}
\begin{figure*}
    \centering   
    \includegraphics[width=0.32\linewidth]{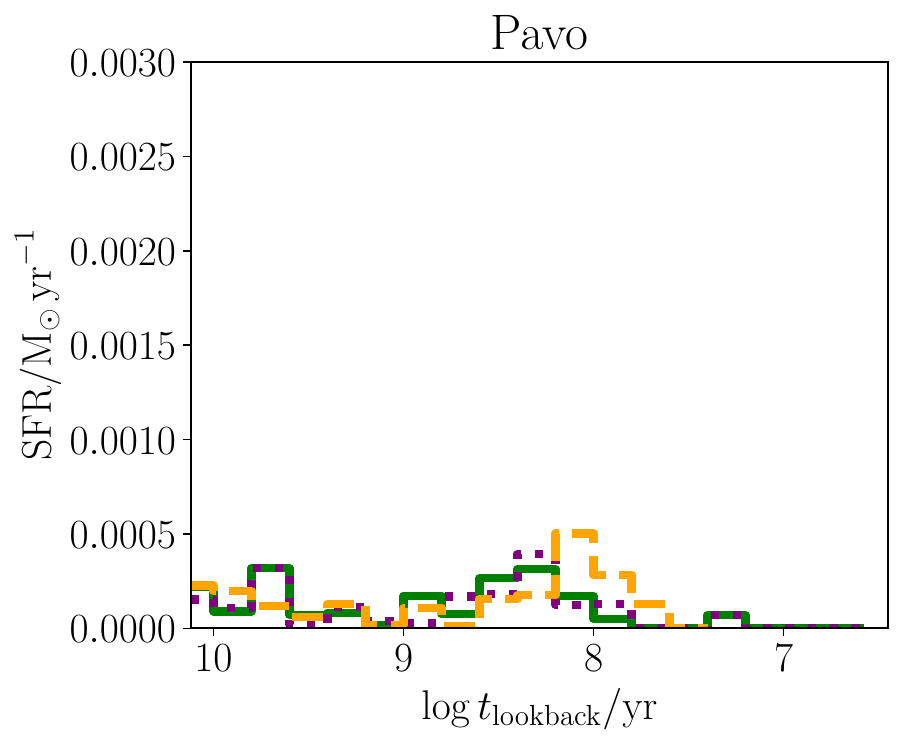}
    \includegraphics[width=0.32\linewidth]{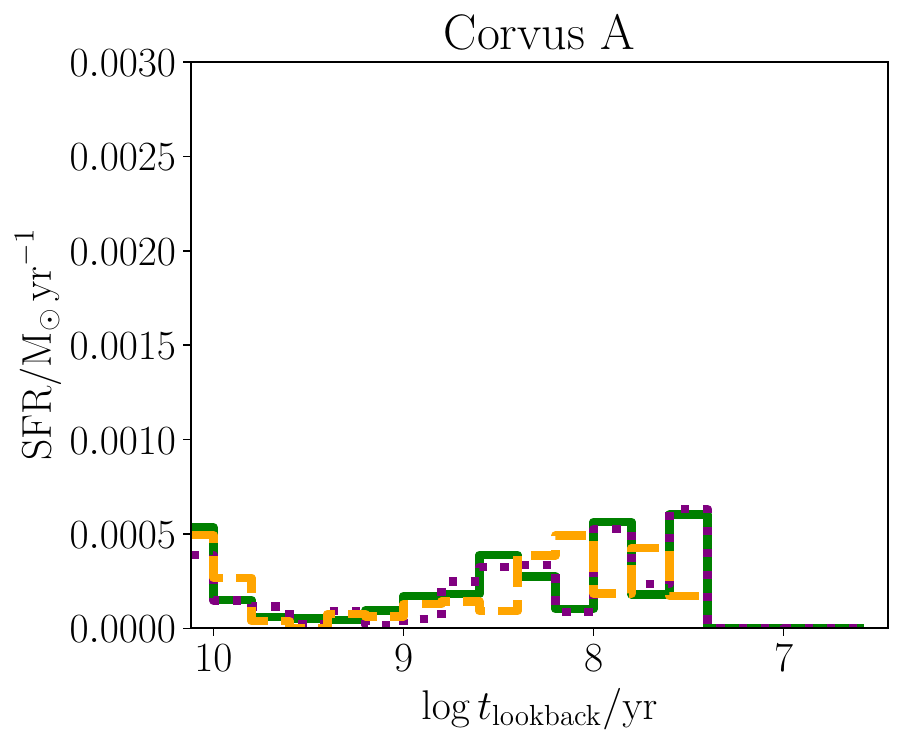}
    \includegraphics[width=0.32\linewidth]{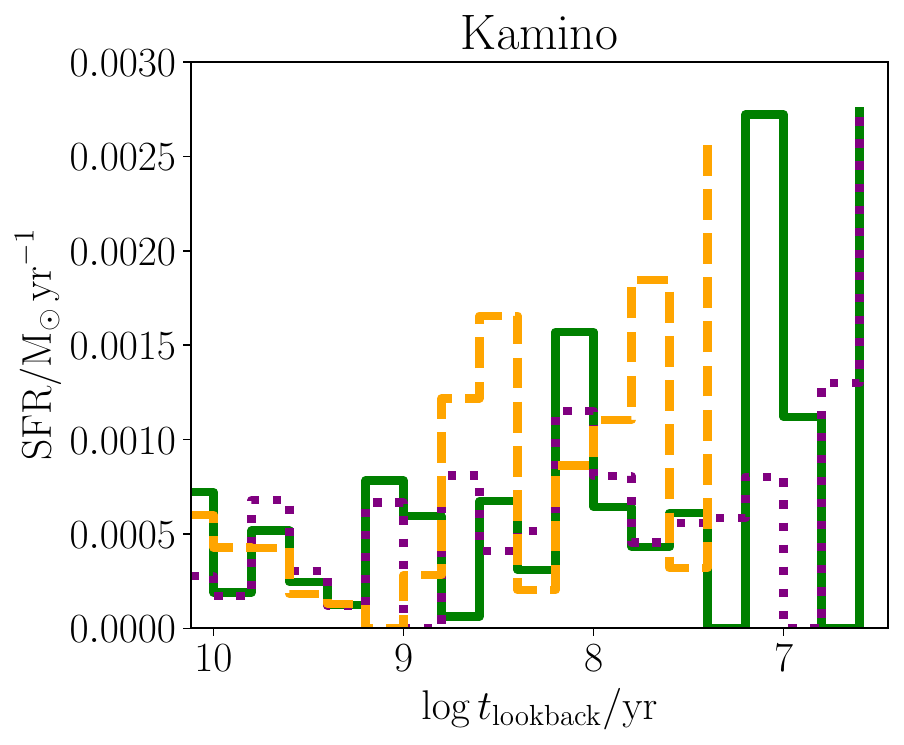}
    \caption{Differential star formation histories of Pavo (left), Corvus~A (center), and Kamino (right).}
    \label{fig:diffSFHs}
\end{figure*}

We select three widely used isochrone sets to fit these SFHs: PARSEC \citep{Bressan2012}, MIST (MESA Isochrones and Stellar Tracks; \citealt{Dotter2016,Choi2016}), and BaSTI (the Bag of Stellar Tracks and Isochrones; \citealt{Pietrinferni2004,Hidalgo2018}). With bins of approximately 0.05~dex in age and 0.1~dex in metallicity, covering the ranges $6.5 < \log t < 10.1$ and $-2.5 < [\mathrm{M/H}] < -1.3$ respectively, we construct artificial binned CMDs for every SSP. These assume a Kroupa \citep{Kroupa+2001,Kroupa+2002} initial mass function (IMF), with minimum and maximum initial masses of 0.1~M$_{\odot}$ \ and 100~M$_{\odot}$, respectively, and a binary fraction of 0.35. The completeness and photometry uncertainties are determined from our artificial star tests. For Pavo and Corvus~A, template contaminant CMDs were generated from the WFC3 parallel fields, while for Kamino we used an ACS field offset from the dwarf, as no WFC3 parallel observations exist for this galaxy. The contribution of random uncertainties is determined by resampling and refitting the best-fit SFH (colored bands in Figure~\ref{fig:cumSFHs}), while the differences between the three isochrone libraries indicate the scale of systematic uncertainties. We note that although our SFHs are computed over a range of metallicity values, our algorithm does not provide reliable constraints on metallicity because metallicity is not treated as a fully free parameter, being limited by the age–metallicity degeneracy in CMD fitting and by constraints imposed to avoid unphysical stellar populations \citep{Mike25_sfh}.

Figure~\ref{fig:cumSFHs} presents the resulting cumulative SFHs, and Figure~\ref{fig:diffSFHs} shows the differential SFHs of the same fits with a logarithmic axis in time. Together with the CMDs in Figure~\ref{fig:cmd}, these figures provide a coherent picture of each dwarf's evolutionary path. All three dwarfs appear to have been continuously building up stellar mass over almost their entire histories. However, while both Pavo and Kamino have had relatively steady build-ups (at least at the temporal resolution of the SFHs), Corvus~A appears to have formed $\sim$60\% of its total stellar mass by 10~Gyr ago. The SFHs of Pavo and Corvus~A further indicate that both systems have recently stopped forming stars (Figure~\ref{fig:diffSFHs}), while Kamino continues to form stars at present. These findings are consistent with the follow-up observations presented in \citet{Jones23,Jones24}, which report UV emission in Swift imaging for both Pavo and Corvus~A. The NUV luminosity traces star formation over the past 100–200~Myr, while H$\alpha$ emission primarily probes the most recent $\sim$10~Myr of activity \citep[e.g.,][]{Lee2009,Kennicutt2012}. The UV detections therefore suggest that star formation persisted until relatively recently -- with average star formation rates of $\sim1\times10^{-4}$ M$_{\odot}$ yr$^{-1}$ for Pavo and $\sim5.6\times10^{-4}$ M$\odot$ yr$^{-1}$ for Corvus~A -- yet neither galaxy shows evidence of \hii\ regions in H$\alpha$ imaging (SOAR for Pavo; Kuiper for Corvus~A), consistent with the SFH-based conclusion that both systems have recently ceased forming stars. In contrast, Kamino exhibits clear H$\alpha$ emission (see Section~\ref{fig:spec}), supporting our SFH result that it is the only system in our sample still forming stars.

Caution is warranted when interpreting the very youngest bins ($<$$100$~Myr) in Figure~\ref{fig:diffSFHs}. In Pavo, for example, the differential SFH formally allows low-level activity down to $\sim$30-60~Myr ago, yet its CMD shows no stars younger than $\sim$150~Myr. Given that Pavo has no \hii~ regions and no H$\alpha$ emission \citep{Jones23}, we favor the CMD-based age limits of $<100$~Myr. By contrast, Kamino exhibits a clear \hii\ region and H$\alpha$ emission (see Section~\ref{fig:spec}), and its CMD contains very young stars (consistent with a 30~Myr isochrone, see Figure~\ref{fig:cmd}), in agreement with ongoing star formation. Because of the stochastic nature of populating the IMF in these systems, the very recent SFHs should be treated with caution. Indicators such as UV flux, H$\alpha$ emission, and the presence of the youngest stars in the CMD provide an independent check on recent star formation activity. If these tracers disagree with the most recent episode inferred from the SFH fitting, greater weight should generally be given to the observational tracers. Similarly, the very early time bins should also be treated with caution, as with the depth of the current photometry, there is considerable scope for degeneracies between ancient SSPs.

\subsection{Stellar Masses \label{sec:mass}}

We determine the present-day stellar mass of Pavo, Corvus~A, and Kamino from their modeled SFHs, following the same approach as \citet{McQuinn2024}, applying a recycling factor of 43\% (which is appropriate for low-metallicity stellar populations with a Kroupa IMF, \citealt{Vincenzo+2016}). This recycling factor represents the fraction of stellar mass returned to the ISM via stellar winds and supernova explosions. We multiply the total stellar mass accumulated over each galaxy's SFH by 0.57, treating recycling as instantaneous. 

The final stellar mass for each dwarf is determined as the median value across all Monte Carlo iterations for the three isochrone sets combined, with uncertainties defined by the central 68\% confidence interval. Because the binary fraction is unknown but affects the absolute normalization of the derived SFHs, we incorporate an additional uncertainty representing a broad range of plausible binary fractions ($0.25 < f_\mathrm{binary} < 0.75$). 

The stellar masses derived for our dwarfs are listed in Table~\ref{tab:dwarfs}: $\log(M_\ast/M_{\odot})$ = $6.08^{+0.14}_{-0.04}$ for Pavo, $6.20^{+0.15}_{-0.07}$ for Corvus~A, and $6.50^{+0.15}_{-0.11}$ for Kamino. Pavo is therefore the lowest-mass dwarf of the three, although all lie just above Leo~P in stellar mass ($\log(M_\ast/M_{\odot})$ = $5.43^{+0.06}_{-0.07}$, \citealt{McQuinn2024}).

\subsection{Luminosities \label{sec:lum}}

To estimate galaxy luminosities, we construct mock stellar populations spanning the full stellar mass range (from 0.1 M$_{\odot}$ up to the most massive surviving star in each SSP) using the Kroupa IMF and corresponding isochrones for every SSP used in the SFH fit. The individual magnitudes of stars in these populations are then combined, weighted according to the derived SFH, to obtain the total magnitudes in each filter (F606W and F814W). Further details of this methodology are presented in \citet{Mike25_sfh}.

We adopt the median magnitude across all Monte Carlo realizations from the three sets of isochrones and quantify uncertainties using the central 68\% confidence interval. We do not include additional uncertainty from binary fraction variations, as its impact on luminosity is minor compared to stellar mass estimates. Finally, the derived F606W and F814W magnitudes are converted to the V-band using the relations from \citet{Sirianni+2005}.

Our derived absolute magnitudes are $M_V=-10.62\pm0.08$ for Pavo, $-10.91\pm0.10$ for Corvus~A, and $-12.02\pm0.12$ for Kamino. For comparison, Leo~P has $M_V=-9.27\pm0.20$ \citep{McQuinn2015}, making it more than one magnitude fainter than any of our dwarfs.

\begin{table*}
\centering
\normalsize
\caption{Properties of our targets, along with the prototypical, isolated field galaxy Leo~P \label{tab:dwarfs}}
\begin{tabular}{lcccc}
\tablewidth{0pt}
\hline
\hline
Parameter & Pavo & Corvus~A & Kamino & Leo~P\\
\hline
R.A. (deg)  & 298.75180$\pm$2\arcsec & 183.69021$\pm$1\arcsec & 63.95506$\pm$2\arcsec & 155.43669$\pm$1\arcsec \\
Dec. (deg) & -61.075240$\pm$1\arcsec & -16.396367$\pm$1\arcsec  & -60.73106$\pm$1\arcsec & 18.089652$\pm$2\arcsec \\
F814W$_{\rm TRGB}$ (mag) & 22.65$\pm$0.05 & 23.61$\pm$0.04  & 24.86$\pm$0.07 & \nodata \\
$m-M$ (mag) & 26.67$\pm$0.08 & 27.62$\pm$0.07  &  28.88$\pm$0.09 & 26.05$\pm$0.20$^{\dagger}$  \\  
$D$ (Mpc)  & 2.16$^{+0.08}_{-0.07}$ & 3.34$^{+0.11}_{-0.11}$  & 5.96$^{+0.26}_{-0.25}$ & 1.62$\pm$0.15$^{\dagger}$ \\ 
$M_{V}$ (mag) & -10.62$\pm$0.08 & -10.91$\pm$0.10   & -12.02$\pm$0.12 & -9.27$\pm$0.20$^{\dagger}$ \\ 
$\log(M_\ast/M_{\odot})$  & 6.08$^{+0.14}_{-0.04}$ & 6.20$^{+0.15}_{-0.07}$   & 6.50$^{+0.15}_{-0.11}$ & 5.43$^{+0.06}_{-0.07}$$^{\ddagger}$  \\
$\log(M_\mathrm{HI}/M_{\odot})$ & 5.79$\pm$0.05 & 6.55$\pm$0.06   & $<$7.3  & 5.91$^{\dagger}$ \\
$v$ (km s$^{-1}$) & 223.7$\pm$3.6$^a$ & 523$\pm$2$^b$ & 526$\pm$21 & 264$\pm$2$^c$  \\
$r_{h}$ (arcsec)     & 49$\pm$2 & 18$\pm$1   & 19$\pm$1 & 46$\pm$2\\     
$r_{h}$ (pc)   	     & 513$\pm$21 & 292$\pm$16  & 506$\pm$27 & 361$\pm$9\\  
$\epsilon$           & 0.52$\pm$0.02 & 0.36$\pm$0.04  & 0.52$\pm$0.03 & 0.65$\pm$0.02 \\ 
Position Angle (deg) & 137$\pm$1 & 117$\pm$4  &  74$\pm$3  & -22$\pm$1 \\

\hline
\end{tabular}
  \begin{tablenotes}
      \small
      \item  R.A.: the Right Ascension (J2000.0). DEC: the Declination (J2000.0). F814W$_{\rm TRGB}$: TRGB magnitude in F814W. $m-M$: the distance modulus. $D$: the distance of the galaxy in Mpc. $M_{V}$: the absolute V-band magnitude. $M_{\rm star}$: the stellar mass in solar mass, derived from SFH (see Section~\ref{sec:sfh}). $log(M_{HI}/M_{\odot})$: the \hi~mass of each object or their $3\sigma$ upper limits. $v$: radial velocities. $r_{h}$: the elliptical half-light radius along the semi-major axis. $\epsilon$: ellipticity which is defined as $\epsilon=1-b/a$, where $b$ is the semiminor axis and $a$ is the semimajor axis. 
      \item \textbf{References:} ($\dagger$) \citet{McQuinn2015}; ($\ddagger$) \citet{McQuinn2024}; ($a$) \citet{Jones25} ; ($b$) \citet{Jones24}; ($c$) \citet{Giovanelli2013}
    \end{tablenotes}
\end{table*}

\subsection{Optical Spectroscopy of Kamino}
\begin{figure}[!b]
\centering
\includegraphics[width = \linewidth]{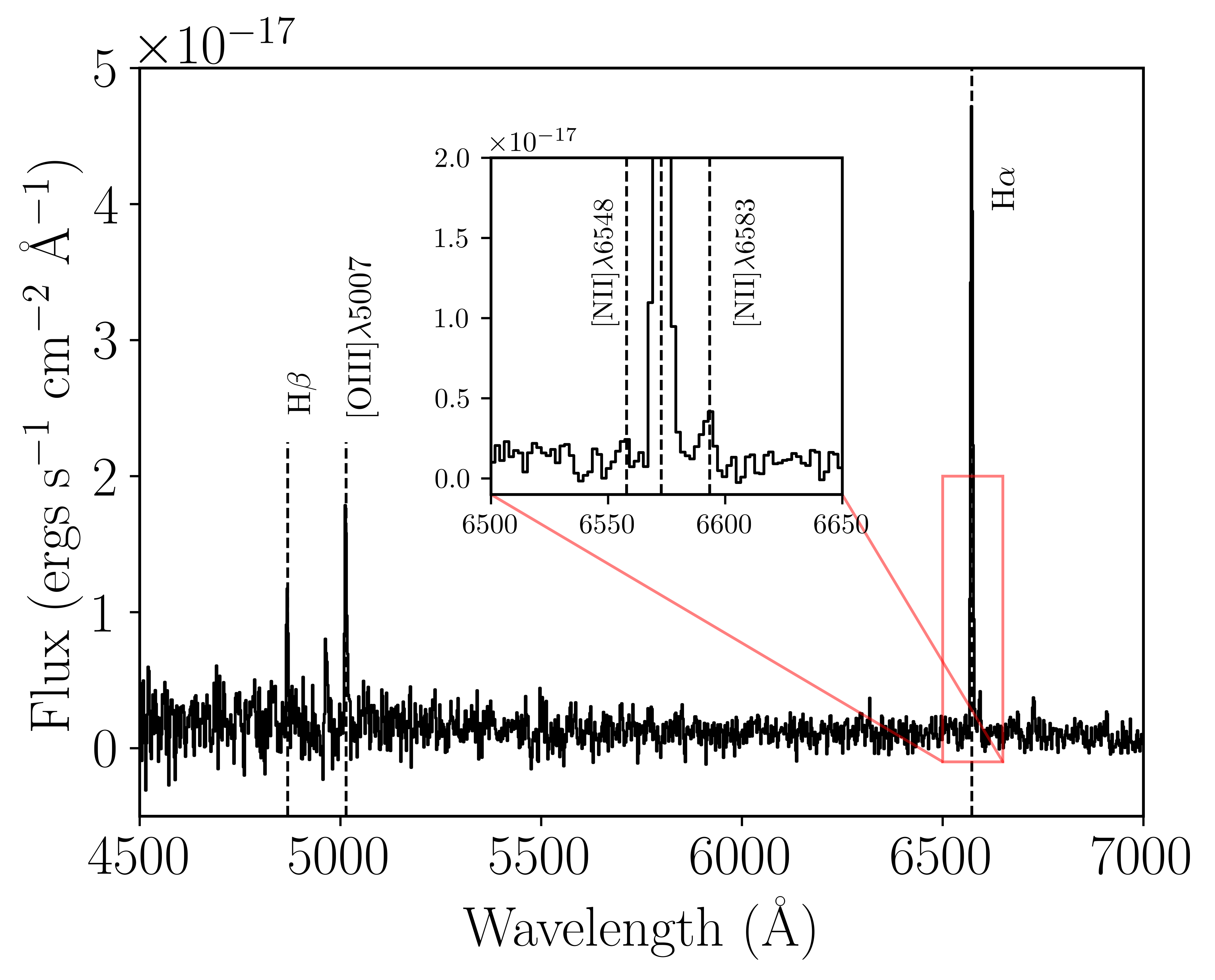}
\caption{The SOAR spectrum of Kamino, with the emission lines used in our analysis highlighted. \label{fig:spec}}
\end{figure}

For Kamino, presented here for the first time, we obtained optical spectroscopy on 2023 December~24 using the SOAR Telescope with the Goodman High Throughput Spectrograph \citep{Clemens2004}. The total exposure time was 45~mins with the 1.2\arcsec\ long slit and 400~lines/mm grating. Data reduction -- including bias-subtraction, flat-fielding, wavelength calibration, and flux calibration -- was carried out using IRAF \citep{Tody1986}. The final reduced spectrum spans a wavelength range of $3999-7860$~\r{A} (see Figure~\ref{fig:spec}).

We extracted the [NII]$\lambda6583$, H$\alpha$, H$\beta$, and the [OIII]$\lambda5007$ lines from a single \hii~region, and measured their fluxes by fitting Gaussians to these spectral features with the \texttt{specutils} package. Extinction corrections were applied following the model of \citet{Gordon2023}. Based on these fits, we derive an H$\alpha$ line velocity of $526\pm21$~km~s$^{-1}$, corrected to the barycentric frame. Using the strong-line calibration relationships from \cite{curti_2020} for the [NII]$\lambda6583$ and H$\alpha$ lines, we derive a metallicity of 12 + log(O/H) = 8.45$\pm$0.09 for Kamino. Applying the two line-ratio diagnostics from \citet{Pettini2004}, we find 12 + log(O/H) = 8.26$\pm$0.18 from N2=[NII]/H$\alpha$, and 8.31$\pm$0.14 from O3N2=([OIII]/H$\beta$)/([NII]/H$\alpha$). Because most strong-line metallicity calibrations deliberately rely on emission-line pairs that are close in wavelength, extinction effects are effectively negligible \citep[e.g.,][]{Poetrodjojo2021}. We emphasize that our approach differs from the analysis of Leo~P in \citet{Skillman2013}, where the metallicity is determined from a direct oxygen abundance measurement rather than a strong-line calibration. For reference, the metallicity of Leo~P is 12 + log(O/H) = 7.17$\pm$0.04. 

Based on the mass–metallicity relation of \citet{Andrews13}, our strong-line metallicity implies a stellar mass of $\log(M_\ast/M_{\odot}) = 8.8 \pm 0.6$, significantly higher than our derived mass of $\log(M_\ast/M_{\odot})= 6.50^{+0.15}_{-0.11}$ (Section~\ref{sec:mass}). This discrepancy likely reflects the breakdown of strong-line calibrations at very low masses \citep[e.g.,][]{Berg2012}, although similar galaxies in the Survey of \hi \ in Extremely Low-mass Dwarfs (SHIELD; \citealt{Cannon2011}) remain consistent with the canonical mass–metallicity relation, unlike Kamino, when using strong-line methods \citep{Haurberg2015}. Another factor may be Kamino's gas content: in a system with little \hi, the injection of metals from recent star formation could drive up the ISM abundance, yielding an anomalously high metallicity. A similar situation may be occurring in Pavo, which has a lower than expected \hi\ mass \citep{Jones25} but no \hii\ region for a gas-phase metallicity measurement. This effect could explain the result in Kamino, where we only have an upper limit on the \hi\ mass (although that upper limit is unfortunately quite large). We further note that Kamino's RGB stars are consistent with a metal-poor stellar population (see Figure~\ref{fig:cmd}), in contrast to the higher metallicity implied by its emission lines. This apparent stellar–gas offset is intriguing but should be interpreted cautiously until more accurate metallicity measurements, such as those based on direct abundance methods, become available.

\subsection{\hi \ masses}

Although Pavo was initially undetected \citep{Jones23} in the \hi \ Parkes All Sky Survey \citep[HIPASS;][]{Barnes2001}, its low mass \hi \ reservoir was recently detected with MeerKAT observations \citep{Jones25}, revealing its \hi \ mass to be $\log(M_{\text{\hi}}/M_{\sun})=5.79$. Corvus~A's \hi \ flux was measured from VLA data by \citet{Jones24} and here we update its \hi \ mass to $\log(M_{\text{\hi}}/M_{\sun})=6.55$, based on our new distance. 

As no targeted \hi \ observations of Kamino currently exist, we extracted a spectrum from HIPASS\footnote{\url{https://www.parkes.atnf.csiro.au/research/multibeam/release/}} at the location of Kamino and searched for any emission corresponding to the H$\alpha$ velocity measurement. Unfortunately, Kamino is in a region of HIPASS with higher than average noise, and no clear signal could be identified. We measured the rms noise in the Hanning smoothed spectrum (velocity resolution of 26~\kms) as 28~mJy. Assuming that Kamino's velocity width is 26~\kms \ or less \citep[for comparison, Leo~P's velocity width is 25~\kms,][]{Giovanelli13}, this equates to a 3$\sigma$ upper limit on its \hi \ mass of $\log M_\mathrm{HI}/M_\odot < 7.3$. Considering Kamino's stellar mass, it is plausible that the galaxy is gas-rich but not detected in HIPASS, similar to Pavo.

\section{Discussion and Conclusions}\label{sec:conclusion}

We have presented new {\it HST} imaging of three newly discovered isolated dwarf galaxies from the SEAMLESS survey -- Pavo, Corvus~A, and Kamino. Our data resolve each dwarf into stars, allowing us to measure their distances, structural properties, and recent star formation histories (SFHs). In this section, we compare our results with those of other isolated dwarf galaxies and Local Group satellites of similar luminosity. We examine their environments, their positions within the size-luminosity space, and their SFHs, and discuss implications for the formation and evolution of isolated dwarf galaxies. 

\begin{figure}[!ht]
\centering
\includegraphics[width = \linewidth]{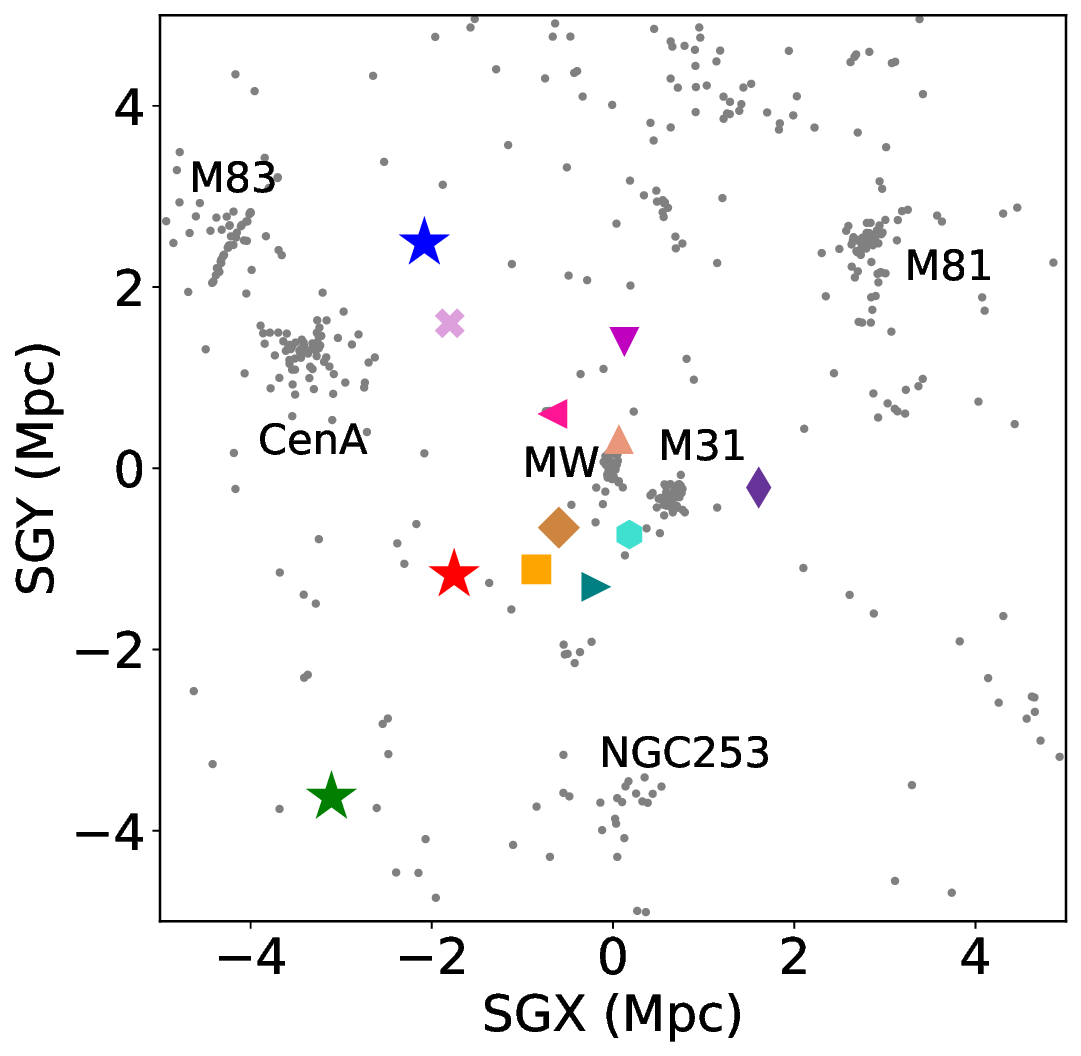}
\includegraphics[width = \linewidth]{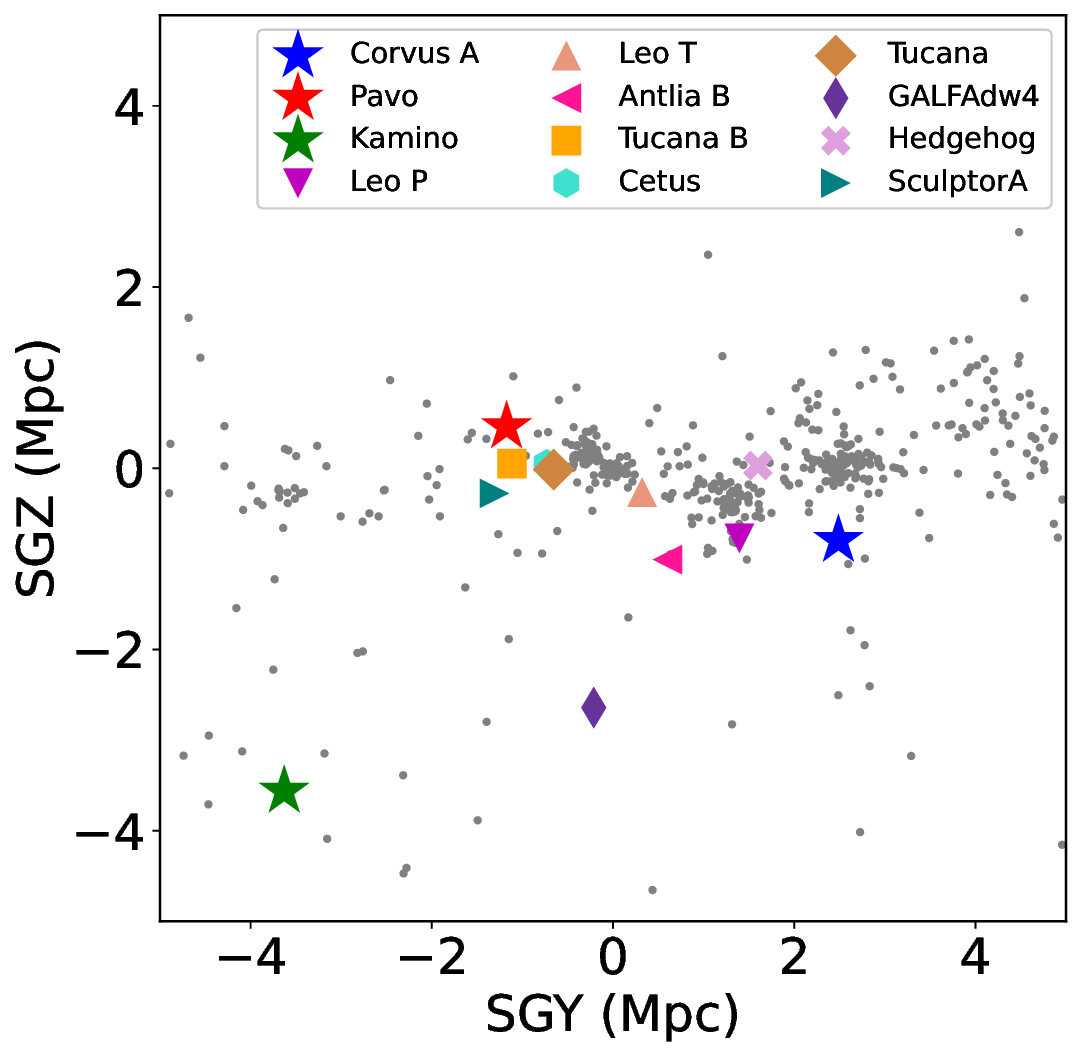}
\caption{A map of nearby galaxies in supergalactic coordinates, illustrating the environment of Pavo, Corvus~A, and Kamino. Top: The supergalactic $X-Y$. Bottom: We show the supergalactic $Y-Z$ plane and the local plane of galaxies. We show several other dwarfs in low-density or isolated environments for reference: Leo~P \citep{Giovanelli13}; Antlia~B \citep{Sand15b}; GALFA-Dw4 \citep{Bennet2022}; Tucana B \citep{Sand2022}; Hedgehog \citep{Li2024}; and Sculptor~A \citep{Sand2024}. Grey points are taken from the compilations of \citep{Karachentsev2004,Karachentsev2019}. We also mark Leo~T, Cetus, and Tucana.  \label{fig:env}}
\end{figure}

To investigate the environment of our targets, we plot their positions (along with other Local Volume dwarfs from \citealt{Karachentsev2019}) in two projections of the supergalactic coordinate system in Figure~\ref{fig:env}. In the plot, we also highlight several other dwarfs in low-density or isolated environments for comparison. As highlighted in discovery papers, Pavo and Corvus~A are part of the local sheet, similar to Leo~P, but in a direction away from any known nearby group/structure. Their nearest neighbors are IC~4662 (3D separation of 721~kpc) and PGC~1059300 (1.0~Mpc), respectively. Kamino is notably distant from the local sheet or any known group. Its closest neighbor, NGC~1705, lies at a separation of 947~kpc -- an isolation comparable to Corvus~A. For further context, GALFA-Dw4 is also located outside the sheet but is significantly more isolated, with its nearest neighbor (HIPASS~J0630$+$08) situated at 2.2~Mpc away from it \citep{Bennet2022}. Hedgehog, by comparison, has its closest neighbor at 980~kpc \citep{Li2024}, similar to Corvus~A and Kamino. These comparisons demonstrate that Pavo, Corvus~A, and Kamino are among the most isolated dwarfs currently known in the Local Volume.

\begin{figure}
\centering
\includegraphics[width = \linewidth]{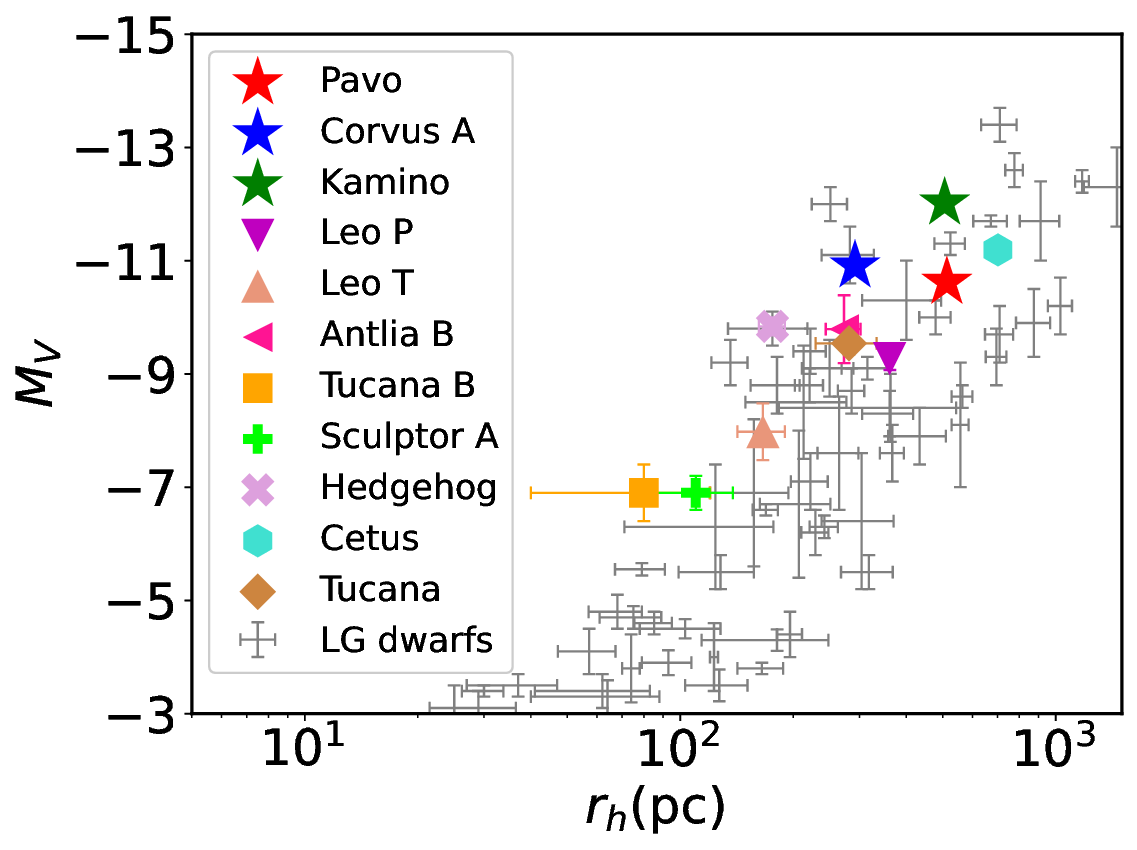}
\caption{Absolute magnitude as a function of half-light radius of Pavo, Corvus~A, and Kamino, relative to Local Volume dwarfs \citep{McConnachie2012,McQuinn2024,Bennet2022,Sand2022,Li2024,Sand2024}. Also plotted are the Local Group dwarf satellites for context \citep{LVDB2024}. For some data points, the error bars are smaller than the symbols and are therefore not visible.
\label{fig:mv-rh}}
\end{figure}

In Figure~\ref{fig:mv-rh}, we compare our dwarfs with Local Group dwarfs and several other notable dwarfs within the Local Volume in the size-luminosity parameter space. All three systems fall within the locus occupied by known dwarfs, with absolute magnitudes $M_V$$\approx$$-9$ to $-12$ and half-light radii of $200-700$~pc, consistent with faint dwarfs in these various environments. Kamino is brighter than Cetus but relatively smaller in size. Pavo and Kamino have comparable physical sizes, though Kamino is $\sim$1.5~mag brighter. Corvus~A is the most compact among our three dwarfs, with a half-light radius similar to those of Antlia~B, Tucana, and Leo~P; however, it is among the brightest galaxies at this size. Notably, the most isolated dwarfs to date preferentially occupy the high-surface-brightness (more compact) side of the size-luminosity relation, reflecting an observational bias in which more compact systems are easier to detect at fixed luminosity. Our ongoing SEAMLESS survey will systematically probe the field dwarf population in the Local Volume. We are also in the process of characterizing our survey completeness, which will shed light on whether the observed isolated systems are representative of the broader population, yielding a better understanding of the true size-luminosity distribution.

\begin{figure}[!b]
\centering
\includegraphics[width =\linewidth]{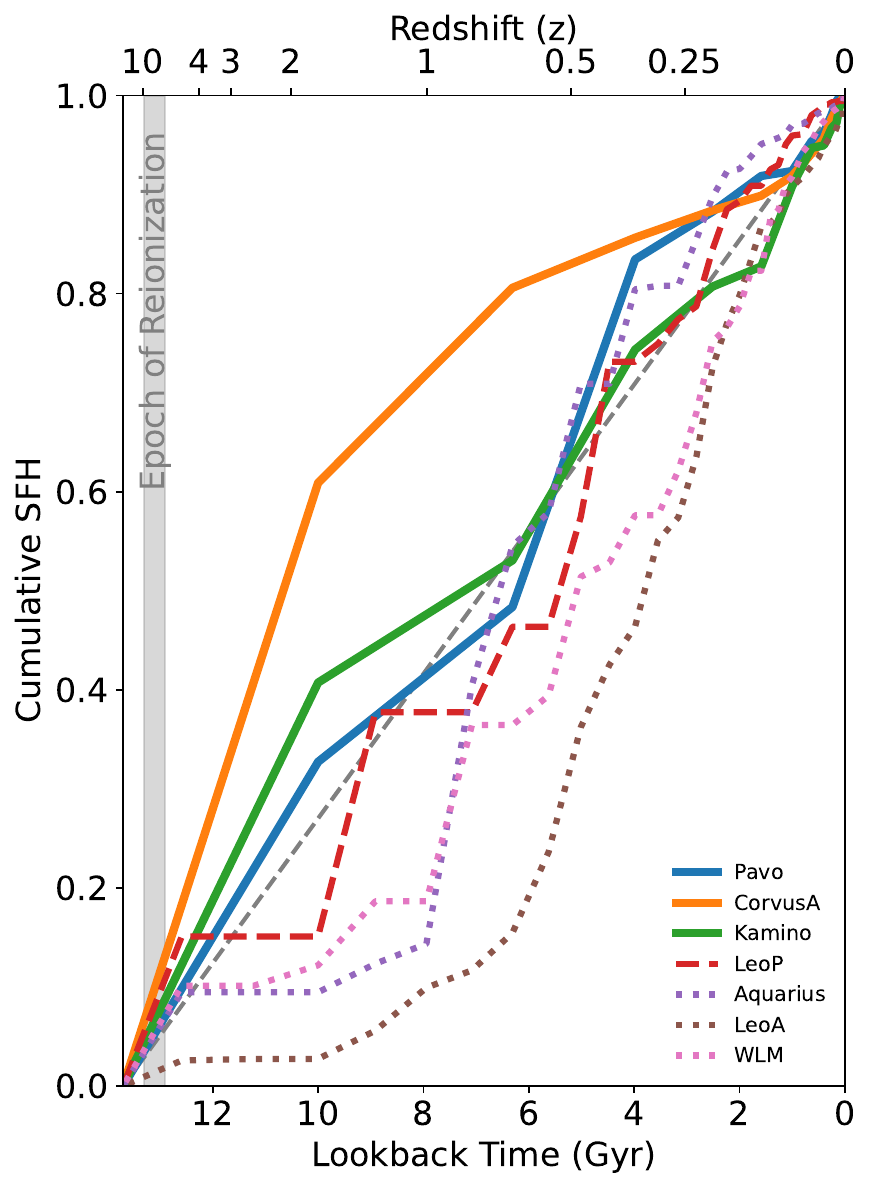}
\caption{The HST-based SFHs of our dwarfs in comparison to the SFHs of the gas-rich isolated low-mass galaxies Leo~P, Aquarius, Leo A, and WLM \citep{McQuinn2024}. The SFHs for Aquarius and Leo A are derived from deeper HST imaging \citep{Cole2007,Cole2014}, and the SFH for WLM is derived from deep JWST NIRCam imaging \citep{McQuinn2024_wlm}. The approximate Epoch of Reionization is shown as a shaded gray bar, and the dashed gray line represents a constant star formation rate across all lookback times. All these SFHs are measured with the PARSEC stellar library and the same Kroupa IMF.
\label{fig:comparison}}
\end{figure}

Finally, in Figure~\ref{fig:comparison}, we compare our cumulative SFHs with those of several well-studied isolated, gas-rich, low-mass galaxies: Leo~P, Aquarius, Leo~A, and WLM \citep{McQuinn2024}. These systems span a similar luminosity range ($M_V\sim-9$ to $-14$) and provide some of the best-resolved SFHs available for isolated dwarfs: Leo~P as a field dwarf beyond the Local Group, and Aquarius, Leo~A, and WLM as relatively isolated systems on the periphery of the Local Group with low probability of previous tidal interactions \citep{McConnachie2012,Shaya2013}. Although these comparison SFHs were derived using different methods, we recovered a similar SFH for Leo~P by applying our methodology to the same HST data used by \citet{McQuinn2015}, which supports the validity of this comparison \citep{Mike25_sfh}. We also note that the SFHs for Leo~A, Aquarius, and WLM are derived from significantly deeper observations, reaching below the old main-sequence turnoff. Leo~A and Aquarius SFHs are based on deep HST/ACS imaging \citep{Cole2007,Cole2014}, whereas WLM SFH is derived from JWST/NIRCam data \citep{McQuinn2024_wlm}. Although our data are shallower than those available for these other galaxies, they nonetheless provide valuable insights into their evolutionary histories until deeper imaging becomes available. 
\citet{McQuinn2024} suggested a characteristic three-stage pattern in the SFHs of isolated low-mass galaxies: an early star formation, a prolonged quiescent interval, and a subsequent reignition of activity. Intriguingly, our targets do not follow this simple pattern. Instead, they exhibit diverse evolutionary pathways, with Corvus~A in particular standing out as highly inconsistent with the common trend seen in the comparison galaxies. This diversity is more consistent with results from observational programs such as Local Cosmology from Isolated Dwarfs (LCID, \citealt{Gallart2015}) and the ACS Nearby Galaxy Survey Treasury \citep{Dalcanton2009} that have shown a rich diversity in the SFHs of isolated dwarf galaxies over cosmic time \citep[e.g.,][]{Cole2007, Cole2014, Monelli2010a,Monelli2010b, Weisz2011, Weisz2014,Skillman2014,Skillman2017}. These findings are also in agreement with the theoretical expectations that the evolution of low-mass galaxies is sensitive to a variety of processes, including inhomogeneous reionization, variable mass assembly histories, and stellar feedback, which imprint strong galaxy-to-galaxy variation in their SFHs \citep[e.g.,][]{Garrison-Kimmel2019,Albers2019,Digby2019,Joshi2021}.

Looking forward, upcoming deep, wide-area imaging from facilities like the Rubin Observatory Legacy Survey of Space and Time \citep{Rubin19}, the Roman Space Telescope \citep{roman2019}, and Euclid \citep{Euclid2025} -- combined with advanced machine learning detection techniques like the one used by our SEAMLESS survey -- will significantly improve our capability to detect and study faint dwarf galaxies at unprecedented distances \citep[e.g.,][]{Rodriguez-Wimberly2019,Mutlupakdil21}. This will provide robust statistical samples across \textit{all} environments, enabling stringent tests of cosmological models on the smallest scales and shedding light on the interplay between environment and star formation in low-mass galaxies.

\begin{acknowledgements}
Based on observations made with the NASA/ESA Hubble Space Telescope, obtained at the Space Telescope Science Institute, which is operated by the Association of Universities for Research in Astronomy, Inc., under NASA contract NAS5-26555. All the HST data used in this paper can be found in MAST. These observations are associated with program \# HST-GO-17514. Support for program \# HST-GO-17514 was provided by NASA through a grant from the Space Telescope Science Institute, which is operated by the Association of Universities for Research in Astronomy, Inc., under NASA contract NAS5-26555. 

BMP acknowledges support from NSF grant AST-2508745. DJS acknowledges support from NSF grants AST-2205863 and 2508746. DC acknowledges support from NSF grant AST-2508747.

Based in part on observations obtained at the Southern Astrophysical Research (SOAR) telescope, which is a joint project of the Ministério da Ciência, Tecnologia e Inovações (MCTI/LNA) do Brasil, the US National Science Foundation’s NOIRLab, the University of North Carolina at Chapel Hill (UNC), and Michigan State University (MSU).

\end{acknowledgements}

\vspace{15mm}
\facilities{HST (ACS/WFC3), SOAR, Parkes}

\software{Astropy \citep{astropy13,astropy18}, The IDL Astronomy User's Library \citep{IDLforever}, DOLPHOT \citep{Dolphin2000}}

\bibliographystyle{aasjournal}
\bibliography{reference}

\end{document}